\documentclass[a4paper, 12pt]{extarticle}
\usepackage[utf8]{inputenc}
\usepackage[margin=1.0in]{geometry}
\usepackage[nottoc]{tocbibind}
\usepackage[style=numeric-comp, sorting=none, backend=bibtex, maxbibnames=99, maxcitenames=2]{biblatex} % Use Biber here
\usepackage{graphicx}

\title{Ferroelectric HfO\textsubscript{2}-ZrO\textsubscript{2} multilayers with reduced wake-up}

\author{\small Barnik Mandal\textsuperscript{*1,3}, Adrian-Marie Philippe\textsuperscript{2}, Nathalie Valle\textsuperscript{2}, Emmanuel Defay\textsuperscript{1,3},\\ \small 
Torsten Granzow\textsuperscript{1}, and Sebastjan Glinsek\textsuperscript{1}.\\ 
\small \textsuperscript{1}Smart Materials Unit, Luxembourg Institute of Science and Technology (LIST), \\
\small 41 Rue de Brill, L-4422 Belvaux, Luxembourg.\\ 
\small \textsuperscript{2}Advanced Analyses and Support Unit, Luxembourg Institute of Science and Technology (LIST), \\
\small 41 Rue de Brill, L-4422 Belvaux, Luxembourg.\\ 
\small\textsuperscript{3}University of Luxembourg, 2 Av. de l'Universite L, \\
\small L-4365, Esch-sur-Alzette, Luxembourg.}

\begin{document}
\maketitle

\hfill \break
\hfill \break
\hfill \break
\hfill \break
\hfill \break
\section*{Corresponding author}
Barnik Mandal (barnik.mandal@list.lu)
\section*{Keywords}
multilayer hafnia zirconia; ferroelectric hafnia zirconia; multilayer thin films;

\clearpage
\section*{Abstract}

Since the discovery of ferroelectricity in HfO\textsubscript{2} thin films, significant research has focused on Zr-doped HfO\textsubscript{2} and solid solution (Hf,Zr)O\textsubscript{2} thin films. Functional properties can be further tuned via multilayering, however,  this approach has not yet been fully explored in HfO\textsubscript{2}-ZrO\textsubscript{2} films. This work demonstrates ferroelectricity in a 50 nm thick, solution-processed HfO\textsubscript{2}-ZrO\textsubscript{2} multilayer film, marking it as the thickest multilayer film to date exhibiting ferroelectric properties. The multilayer structure was confirmed through transmission electron microscopy (TEM) and energy dispersive x-ray spectroscopy, with high-resolution TEM revealing grain continuity across multiple layers. This finding indicates that a polar phase in the originally paraelectric ZrO\textsubscript{2} layer, can be stabilized by the HfO\textsubscript{2} layer. The film attains a remanent polarization of 9 µC/cm\textsuperscript{2} and exhibits accelerated wake-up behavior, attributed to its higher breakdown strength resulting from the incorporation of multiple interfaces. These results offer a faster wake-up mechanism for thick ferroelectric hafnia films.

\clearpage
\section{Introduction}

The discovery of ferroelectricity in doped HfO\textsubscript{2} has garnered interest due to its compatibility with complementary metal-oxide-semiconductor (CMOS) technology.\cite{Böscke} Given the similar chemical and physical properties of HfO\textsubscript{2} and ZrO\textsubscript{2}, studies on ferroelectricity in ZrO\textsubscript{2} have also been conducted.\cite{starschich_ZrO2_2017} Both of these simple oxides are regarded as promising lead-free alternatives for non-volatile memory and piezoelectric applications. In 2016, Fan et al. demonstrated the stabilization of the ferroelectric orthorhombic phase
in a ZrO\textsubscript{2} thin film produced by radio frequency magnetron sputtering, utilizing substrate-induced strain to
promote transition from the paraelectric tetragonal phase (t-phase) to the ferroelectric orthorhombic phase (o-phase).\cite{Fan2016} Starschich et al. reported thicker ZrO\textsubscript{2} and Hf-doped ZrO\textsubscript{2} ferroelectric films of 100 nm and 390 nm, respectively, using chemical solution deposition (CSD) with organometallic precursors.\cite{starschich_ZrO2_2017} The evolution of ferroelectricity in solid solutions of HfO\textsubscript{2} and ZrO\textsubscript{2} (Hf\textsubscript{x}Zr\textsubscript{1-x}O\textsubscript{2}) has been intensively studied using various growth techniques, with reported properties ranging from ferroelectric- to antiferroelectric-like. The maximum remanent polarization \textit{P}\textsubscript{r} is observed typically around the composition Hf\textsubscript{0.5}Zr\textsubscript{0.5}O\textsubscript{2}, while existence of a CMOS-compatible morphotropic phase boundary has been postulated for the composition Hf\textsubscript{0.3}Zr\textsubscript{0.7}O\textsubscript{2}.\cite{Nakayama2018, Mller2012, Ni2019}\par

Several experimental investigations reported ferroic HfO\textsubscript{2}-ZrO\textsubscript{2} multilayers.\cite{Lu2016, Chen2022, Weeks2017, Park2019_nanolaminates,Cheema2022} The main drive behind the majority of these studies has been the stabilization of the ferroelectric phase and enhancement of properties in sub-20 nm multilayer films through mechanical confinement of the layers\cite{Lu2016}, increased interface energy\cite{Chen2022} and lattice distortion caused by grain coalescence during thermal annealing.\cite{Weeks2017}  Using Density Functional Theory, Dutta et al. demonstrated that small-size dopants like Si tend to form distinct layers in the orthorhombic ferroelectric polymorph, avoiding intermixing with Hf. Layering stabilizes the polar phase over non-polar polymorphs, strongly suggesting the use of multilayer structures instead of solid solutions.\cite{Dutta2020} Furthermore, Cheema et al. reported permittivity enhancement in ultrathin HfO\textsubscript{2}-ZrO\textsubscript{2} superlattices with mixed ferroelectric–antiferroelectric order, i.e., an effect not possible in conventional ferroelectrics.\cite{Cheema2022} These results demonstrate the potential advantages of multilayers in polar fluorite oxides.

In this work, we examine the stabilization of ferroelectricity and the wake-up behaviour  in 50 nm multilayer La$:$HfO\textsubscript{2}-ZrO\textsubscript{2} thin films, representing the thickest multilayer thin films studied to date. While there have been studies on thick Hf\textsubscript{0.5}Zr\textsubscript{0.5}O\textsubscript{2} solid-solution films,\cite{starschich_ZrO2_2017, Schenk2019} they have not specifically focused on multilayer structures. We prepared pure ZrO\textsubscript{2}, La$:$HfO\textsubscript{2}, and multilayer La$:$HfO\textsubscript{2}-ZrO\textsubscript{2} films using a CSD process described in our previous publication.\cite{Mandal2024} La doping for HfO\textsubscript{2} layer was chosen based on density functional theory predictions, which suggested that La's larger ionic radius and lower electronegativity promote stabilization of the ferroelectric orthorhombic Pca2\textsubscript{1} phase.\cite{Batra2017, Materlik2018} Experimental findings further validated the predictions, demonstrating a wide doping range and reduced polarization relaxation.\cite{Schroeder2018} For clarity in the following discussion, the thin-film samples of La$:$HfO\textsubscript{2}, ZrO\textsubscript{2}, and La$:$HfO\textsubscript{2}-ZrO\textsubscript{2} multilayers are referred to as HO, ZO, and HZO, respectively. It is observed that although the pure ZrO\textsubscript{2} film is paraelectric, the combination with the polar HfO\textsubscript{2} layer induces ferroelectricity in ZrO\textsubscript{2} layer of the HZO film.
High-resolution transmission electron microscopy (HR-TEM) confirms that the polar phase extends through multiple layers.  The HZO film exhibits ferroelectric properties with a remanent polarization (\textit{P}\textsubscript{r}) of 9 µC/cm\textsuperscript{2}, which is minor improvement to the conventional HO films. We further show that the multilayered HZO films can sustain higher electric fields than the HO film alone, resulting in significant reduction of wake-up cycles. The number of cycles required for polarization saturation in HZO decreased tenfold (from 10,000 to 1,000) compared to standard HO films.

\section{Methods}
\emph{Solution Preparation}$:$ Two 0.25 M precursor solutions were prepared, one containing La:HfO\textsubscript{2} for the HO film and the other containing ZrO\textsubscript{2} for the ZO film. Additionally, two solutions of La:HfO\textsubscript{2} and ZrO\textsubscript{2} with a concentration of 0.08 M were prepared for the multilayer HZO film. For 5\% La-doped HfO\textsubscript{2}, Hf(IV)-acetylacetonate (Alfa Aesar, 97\%) and La(III)-acetate hydrate (Sigma-Aldrich, 99.9\%) were used as metal precursors, with the La(III)-acetate hydrate being freeze-dried for 16 hours to remove crystal water. For pure ZrO\textsubscript{2}, Zr(IV)-acetylacetonate (Sigma-Aldrich, 97\%) was used. The powders were dissolved in propionic acid (Sigma-Aldrich, 99.5\% purity) and then refluxed for 3 hours at 150 °C with magnetic stirring in an Ar atmosphere using a modified Schlenk apparatus.
\\ \\
\emph{Film Preparation}$:$ The solutions were spin-coated onto platinized Si substrates (Pt–Si, SINTEF) at 3000 rpm for 30 seconds. The platinum layer was strongly oriented in the (111) direction. After each spin-coating, a drying step was performed on a hot plate at 215 °C for 5 minutes. For the ZO and HO films, three 15 nm-thick amorphous layers were spin-coated from 0.25 M ZrO\textsubscript{2} and 0.25 M La$:$HfO\textsubscript{2} solutions, respectively. For the HZO sample, ten 5 nm-thick layers of La$:$HfO\textsubscript{2} and ZrO\textsubscript{2} were alternately deposited using 0.08 M La$:$HfO\textsubscript{2} and 0.08 M ZrO\textsubscript{2} solutions (see Figure \ref{fig:eds+xrd}(a)). The films were subsequently subjected to rapid thermal annealing (RTA) using an AS-Master 2000 (Annealsys) tool. In the RTA process, the chamber underwent a purging sequence by pumping down to a pressure below 4 mbar. Subsequently, the chamber was restored to atmospheric pressure while being filled with N\textsubscript{2} and O\textsubscript{2}. N\textsubscript{2} and O\textsubscript{2} were pumped at flow rates of 1000 sccm each to establish an atmosphere with a 1$:$1 ratio. Following the chamber filling, the final crystallization step was initiated by reducing gas flow rates to 150 sccm$:$150 sccm for N\textsubscript{2}$:$O\textsubscript{2}. Films were then crystallized at 800 \textdegree C for 90 s with 50 \textdegree C s\textsuperscript{-1} ramping rate. Circular top electrodes were evaporated on the film, each measuring 100 \textmu m in diameter. These electrodes comprised two layers: first a 5 nm thick Ti adhesion layer, followed by a 100 nm thick Pt layer. The deposition was carried out using a Plassys evaporator under a pressure of 5x10\textsuperscript{-8} mbar. The electrodes were patterned by standard lithography and lift-off processes. 
\\ \\
\emph{Characterization}$:$ A Bruker D8 Discover X-ray diffractometer with Cu K$\alpha$ radiation ($\lambda$ = 0.154 nm) was used for grazing incidence (GI) and $\theta$-2$\theta$ X-ray diffraction (XRD). An incidence angle of 0.5\textdegree~ was used for GIXRD. The reference patterns for monoclinic, orthorhombic, and tetragonal HfO\textsubscript{2} were taken from the powder diffraction files with the numbers 00-034-0104, 04-005-5597, and 04-003-2612, respectively.\cite{ICDD_PDF4_plus_v4.19} These same references were also applied to ZrO\textsubscript{2}, as they have very similar crystalline structures.

Ferroelectric measurements were performed using a TF Analyzer 2000 (aixACCT, Germany). Polarization versus electric field loops were acquired using a 3 kHz bipolar triangular signal, following the application of a desired number of rectangular wake-up cycles at 3 kHz. Two distinct sets of wake-up cycles were conducted on the HZO and HO films. The first set, referred to as procedure 1, consisted of a 1000-cycle wake-up at  15 V. The second set, procedure 2, involved 7000 cycles, performed as seven consecutive 1000-cycle wake-ups, with increasing voltages from 8 V to 15 V.

The (Scanning) Transmission Electron Microscopy ((S)TEM) analyses were performed on a JEOL JEM-F200 cold FEG microscope operating at an acceleration voltage of 200 kV. TEM lamella was prepared following the “lift-out” method with a FEI Helios Nanolab 650 Focussed Ion Beam Scanning Electron Microscope (FIB-SEM).  X-ray Energy Dispersive Spectroscopy (EDS), using dual JEOL 100 mm\textsuperscript{2} Silicon Drift Detectors, was performed in STEM mode allowing elemental maps analysis. High Resolution TEM (HRTEM) imaging combined with Fast Fourier Transform (FFT) computation was performed to identify crystalline phases observed throughout the thin film. The reference patterns for monoclinic, orthorhombic, and tetragonal HfO\textsubscript{2} were taken from the powder diffraction files with the numbers 04-002-2772, 04-005-5597, and 04-027-1007, respectively.\cite{ICDD_PDF4_plus_v4.19}

\section{Results and discussion}

\begin{figure}[!ht]
    \centering
        \includegraphics[scale=0.7]{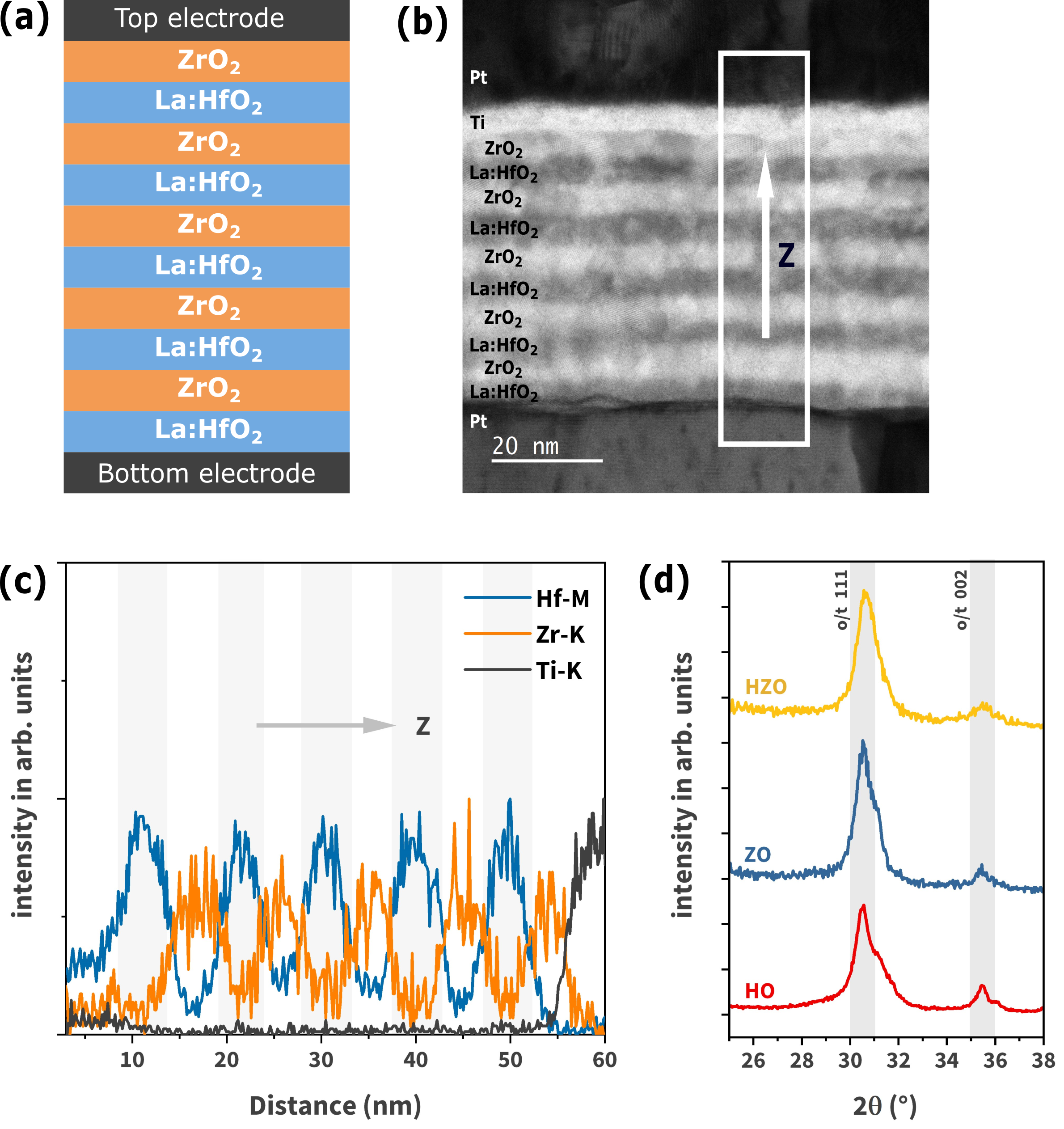}
        \caption{\footnotesize (a) Structural schematics, (b) bright-field  STEM image of HZO multilayer film with Pt/Ti top electrode and Pt bottom electrode, (c) STEM-EDS line profile of the HZO multilayer. Z is the direction of growth. (d) GIXRD patterns of the HO film, ZO film and HZO multilayer film. The main peaks are labeled with o/t (orthorhombic/tetragonal), which correspond to the reference patterns of Pca2\textsubscript{1} and P4\textsubscript{2}/nmc phases, respectively.\cite{ICDD_PDF4_plus_v4.19}}
        \label{fig:eds+xrd}
    \end{figure}

    Figure \ref{fig:eds+xrd}(a) presents a structural schematic of the multilayer HZO film, which consists of ten 5 nm-thick layers of alternately deposited La$:$HfO\textsubscript{2} and ZrO\textsubscript{2}, starting with La:HfO\textsubscript{2} and ending with ZrO\textsubscript{2}. The bright-field STEM image in Figure \ref{fig:eds+xrd}(b) reveals the expected multilayered stack, indicated by the contrast variation between the La:HfO\textsubscript{2} and ZrO\textsubscript{2} layers, with two Pt-based electrodes enclosing the stack. EDS spectrum (see Supporting Information) to generate the line profile in Figure \ref{fig:eds+xrd}(c) was extracted from the area outlined by the white box in Figure \ref{fig:eds+xrd}(b). The oscillating behavior of the Hf M and Zr K EDS line profiles in Figure \ref{fig:eds+xrd}(c), along with EDS cross-section mapping (see Supporting Information) further validates the multilayered structure of the film, with the Ti K line indicating where the film terminates. The grazing incident X-ray diffraction (GIXRD) patterns of the HO, ZO, and HZO films are illustrated in Figure \ref{fig:eds+xrd}(d). The patterns show the characteristic features typically observed in HfO\textsubscript{2} and ZrO\textsubscript{2} thin films, i.e., a pronounced peak at 30.6 ° and a weaker peak at 35.5 °, indicative of the orthorhombic polar planes (111) and (002), respectively. However, note that the structural similarities between the polar orthorhombic and non-polar tetragonal phases lead to significant overlap of their reflections, making it challenging to distinguish them in laboratory XRD experiment. Additionally, a broad shoulder peak between $\sim$31° and 31.5° is also present, which may correspond to a monoclinic phase (P2\textsubscript{1}/c).
    
    \clearpage

\begin{figure}[!ht]
    \centering
        \includegraphics[scale=0.45]{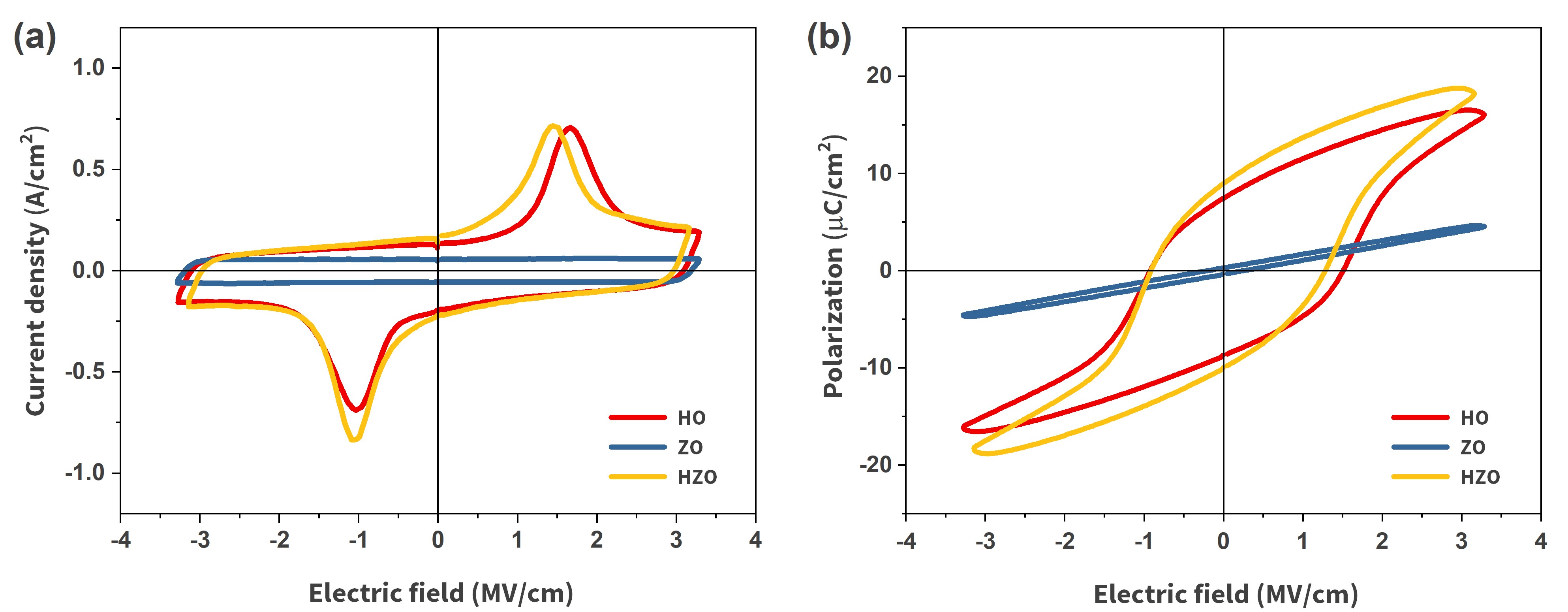}
        \caption{\footnotesize (a) Current density versus electric field loops, and (b) polarization versus electric field loops of the three films. }
    \label{fig:loops}
    \end{figure}

    The current density-electric field and polarization-electric field hysteresis loops after wake-up cycling are shown in Figure \ref{fig:loops}. The ZO film does not show any switching current and is therefore paraelectric. The HO and multilayered HZO films exhibit prominent switching current peaks. The HO film has a positive remanent polarization (\textit{P}\textsubscript{r}) of 8 µC/cm\textsuperscript{2}, with coercive field (\textit{E}\textsubscript{c}) of 1.5 MV cm\textsuperscript{-1}. HZO has a positive \textit{P}\textsubscript{r} of 9 µC/cm\textsuperscript{2}, with E\textsubscript{c} of 1.2 MV cm\textsuperscript{-1}. If the ZrO\textsubscript{2} layers were purely dielectric, they would have acted as dead layers, reducing the effective electric field across the film. This, in turn, would result in a decrease in remanent polarization and an increase in the coercive field\cite{Zhang2013}, which is not observed in this case. Therefore, it can be concluded that HZO and HO samples have effective ferroelectric layers of almost identical thicknesses. Interesting to note is that when multilayers are prepared with ZrO\textsubscript{2} as the initial layer, poor ferroelectric properties are  observed in the multilayer (see Supporting Information). This is likely due to the non-polar phase of the initial ZrO\textsubscript{2} layer. We can speculate that this could be improved if the polar phase would be stabilized in this first layer by optimizing the processing parameters. Similar result is reported by van Gent et al.\cite{Johanna2024} in epitaxial rhombohedral Hf\textsubscript{1-x}Zr\textsubscript{x}-ZrO\textsubscript{2} superlattices prepared via pulsed laser deposition, indicating that this effect is intrinsic to HfO\textsubscript{2}-ZrO\textsubscript{2} multilayers, regardless of the fabrication method, type of ferroelectric phase or microstructure.
    
    \clearpage

\begin{figure}[!ht]
    \centering
        \includegraphics[scale=0.8]{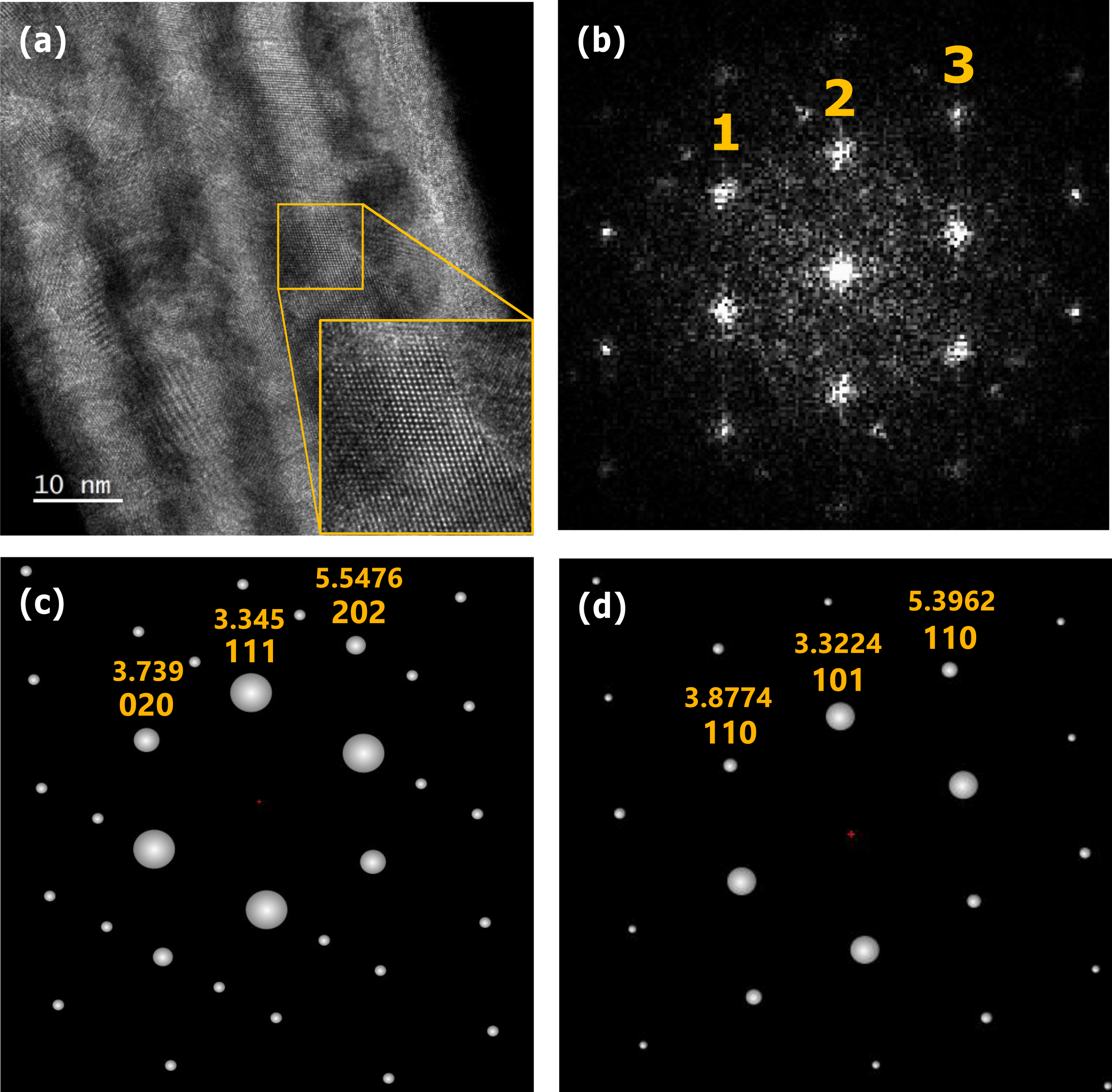}
        \caption{\footnotesize (a) Cross-sectional HRTEM images of field cycled HZO multilayer film. (b) FFT diffractogram of the squared area. Simulated diffraction pattern of (c) orthorhombic HfO\textsubscript{2} along [-101] zone axis (space group Pca2\textsubscript{1}) and (d) tetragonal HfO\textsubscript{2} along [-111] zone axis (space group P4\textsubscript{2}/nmc),  .}
    \label{fig:tem}
    \end{figure}

    To gain a better understanding of the phase and grain orientation, cross-sectional high-resolution transmission electron microscopy (HRTEM) was performed on the HZO sample and the result is shown in Figure \ref{fig:tem}(a). Lattice fringes extending across multiple layers are observed, suggesting a continuation of grains and their orientation through the layers with different composition. The Fast Fourier Transform (FFT) of an exemplary area is presented in Figure \ref{fig:tem}(b). The FFT experimental pattern was compared with the simulated diffraction patterns from PDF cards of monoclinic (space group P2\textsubscript{1}/c), orthorhombic (space group Pca2\textsubscript{1}) and tetragonal phase (space group P4\textsubscript{2}/nmc). It aligns well with the simulated diffraction pattern of the ferroelectric o-phase along the [-101] zone axis (Figure \ref{fig:tem}(c)) and non-polar t-phase along the [-111] zone axis (Figure \ref{fig:tem}(d)). The ratio of the distances from the center to spots 1 and 2, and ratio from the center to spots 2 and 3 in the experimental FFT, are 1.62 and 0.86, respectively. These values are closely aligned with those obtained from simulated patterns of the o-phase and t-phase (see Supporting Information). The observed ferroelectric switching in the electrical measurements suggests the presence of the o-phase rather than the t-phase in the presented lattice fringe. However, the possibility of the diffractogram corresponding to the m-phase can be ruled out (see Supporting Information).

    \clearpage

\begin{figure}[!ht]
    \centering
        \includegraphics[scale=0.5]{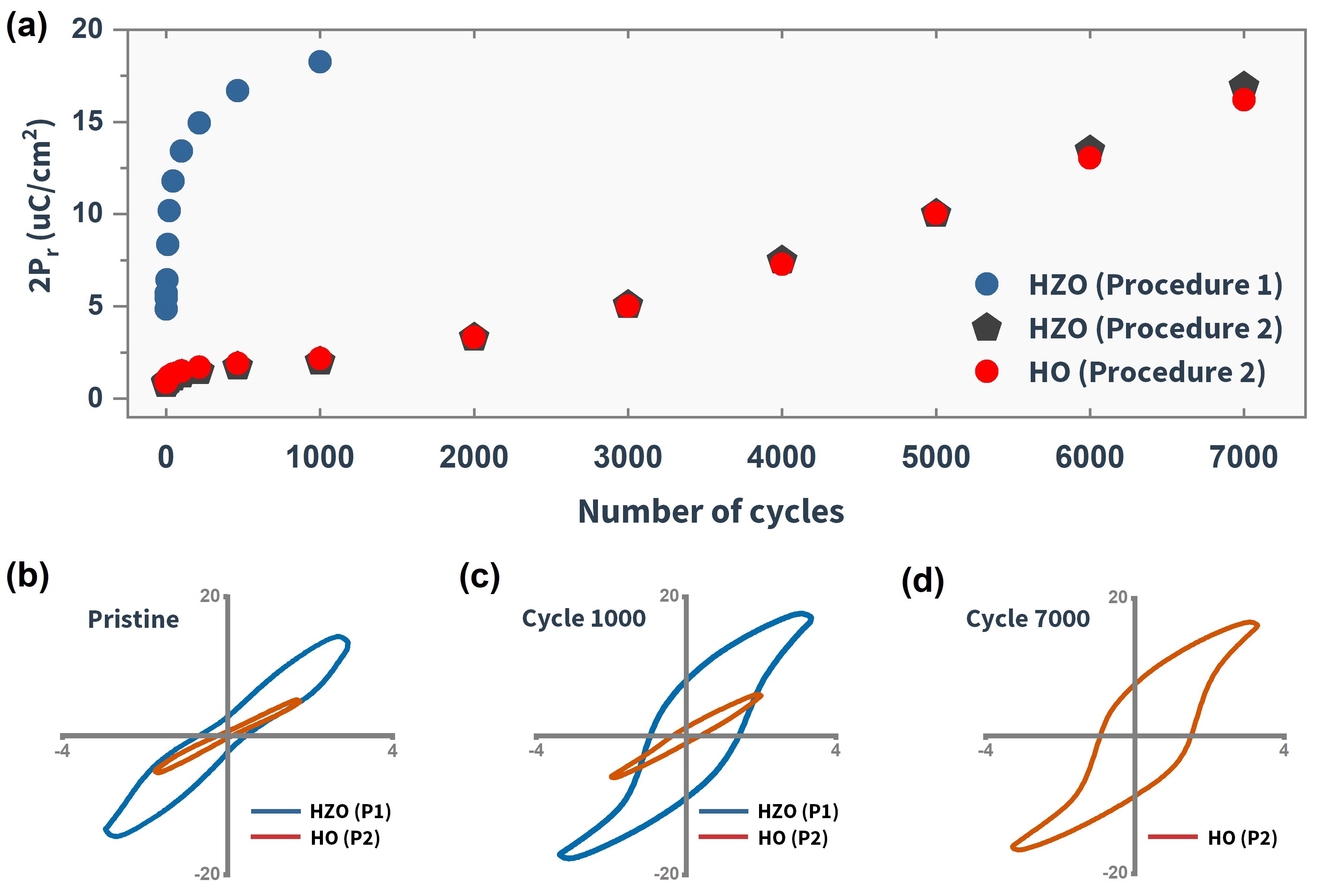}
        \caption{\footnotesize (a) Remanent polarization 2\textit{P}r vs. number of wake-up cycles for HZO and HO films. (b) Pristine loops of HZO and HO, acquired from procedure 1 and procedure 2, respectively. (c) HZO and HO loops measured after 1000 cycles following procedure 1 and procedure 2, respectively. (d) HO loop measured after 7000 cycles following procedure 2, for the HZO film, maximum 2\textit{P}r was achieved by procedure 1 (1000 cycles), therefore no loop is recorded after 7000 cycles. [P1: Procedure 1, P2: Procedure 2]}
    \label{fig:wakeup}
    \end{figure}

    While the conventional HO and multilayer HZO films show comparable ferroelectric properties, an important difference was observed during the wake-up cycling. Figure \ref{fig:wakeup}(a) illustrates the increase of remanent polarization 2\textit{P}\textsubscript{r} with respect to the number of wake-up cycles. Two distinct sets of wake-up cycles were conducted on the HZO and HO films. The first set, referred to as procedure 1, consisted of a 1000-cycle wake-up at an electric field of approximately 3 MV cm\textsuperscript{-1}, exceeding the coercive field of typical HfO\textsubscript{2} films. The second set, procedure 2, involved 7000 cycles, performed as seven consecutive 1000-cycle wake-ups, with the electric field gradually increasing from about 0.8 MV cm\textsuperscript{-1} to around 3 MV cm\textsuperscript{-1}. The HO film could not withstand the immediate application of fields exceeding \textit{E\textsubscript{c}} and typically broke down after approximately 10 cycles, making wake-up cycling using procedure 1 impossible. HZO film, on the other hand, revealed an escalated wake-up behaviour upon treating with procedure 1. The pristine loops of HZO (from procedure 1) and HO (from procedure 2) films are presented in \ref{fig:wakeup}(b), where HZO shows a propeller-shaped hysteresis loop with non-zero remanent polarization, very similar to Hf\textsubscript{x}Zr\textsubscript{1-x}O\textsubscript{2} films reported in previous studies.\cite{Lu2016, Mller2012} Experimental observations and theoretical modeling have linked that to an electric field-induced phase transition from the tetragonal (T) phase to the orthorhombic o-III phase.\cite{ReyesLillo2014, starschich_ZrO2_2017} Furthermore, studies have shown that propeller-shaped hysteresis loops arise due to defect pinning and are also influenced by the evolution of charged domain walls.\cite{Mehmood2020,Zhou2022} Figure \ref{fig:wakeup}(c) presents the hysteresis loops after 1000 cycles, showing that HZO reached a maximum 2\textit{P}\textsubscript{r} of 18 µC/cm\textsuperscript{2}. In comparison, the HO film required 7000 cycles to achieve comparable 2\textit{P}\textsubscript{r} (see Figure \ref{fig:wakeup}(d)). When using procedure 2, HZO showed nearly identical wake-up behavior to the HO sample (see Figure \ref{fig:wakeup}(a)). 
   
    Wake-up field cycling is typically performed in hafnia-based ferroelectrics to improve ferroelectric response through redistribution of defects, typically oxygen vacancies.\cite{Pesic2016, Jiang2020} However, these defects, often located at grain boundaries, interfaces or domain walls, can also form conductive filaments upon exposure to electric field.\cite{Max2018, Xu2024, Lee2023} Performing wake-up cycles directly at or above the coercive field is therefore detrimental to dielectric properties of conventional HO films. We hypothesize that the reason multilayer HZO films can endure direct application of high fields are multiple La$:$HfO\textsubscript{2}-ZrO\textsubscript{2} interfaces, which disrupt the formation of conductive filaments along the film thickness. Breakdown measurements were performed on both samples (see Supporting Information), and early breakdown was indeed observed in the HO film. Furthermore, endurance improved from 10\textsuperscript{4} cycles in HO films to 10\textsuperscript{5} cycles in HZO films (see Supporting Information). Note that enhanced cycling endurance from 10\textsuperscript{6} to 10\textsuperscript{9} cycles is observed in epitaxial rhombohedral Hf\textsubscript{1-x}Zr\textsubscript{x}-ZrO\textsubscript{2} superlattices, confirming that multilayering is an effective approach to robust ferroelectric hafnia-based films.\cite{Johanna2024} Beyond hafnia, Sun et al.\cite{Sun2016} achieved a breakdown strength of 4.5 MV cm\textsuperscript{-1} in BZT-BCT multilayer films, compared to 1 MV cm\textsuperscript{-1} in single-layer BZT and BCT films of the same thickness. They explained the effect by inhibited propagation and growth of "electric trees" in the films with engineered interfaces. In addition to multilayering, early fatigue in ferroelectric thin films has been reported to be tackled through strategies such as engineering of oxygen vacancies, suppressing defect diffusion, and introducing compositional inhomogeneity. \cite{Bao2023, Kang2024, Yan2024}

    \clearpage

\clearpage
\section{Conclusion and outlook}

In summary, we demonstrated ferroelectricity in solution-processed 50 nm-thick HZO multilayer film. The ferroelectric properties of the multilayer thin film were investigated through the characterization of compositional profiles, crystalline phases, and electrical measurements. Although the pure ZrO\textsubscript{2} films processed were paraelectric in nature, the synthesis of a multilayer structure with HfO\textsubscript{2} induced a polar phase in ZrO\textsubscript{2} through the La-doped HfO\textsubscript{2} layer. The HZO film exhibits ferroelectric properties, with a remanent polarization of 9 µC cm\textsuperscript{-2} and a coercive field of 1.2 MV cm\textsuperscript{-1}. Additionally, as prepared HZO films can endure higher electric fields due to the presence of multiple interfaces, which significantly reduces the wake-up cycles - an improvement by an order of magnitude compared to the conventional films. This study highlights a novel pathway to enhance the wake-up properties of ferroelectric HfO\textsubscript{2} and ZrO\textsubscript{2}-based thin films by introducing interfaces, rather than relying solely on solid solutions. This approach can be extended to the growth of thicker multilayer films for piezoelectric applications, as the improved wake-up properties and reduced early breakdown during extended cycling help address the challenges typically associated with thicker films.

\clearpage

\section*{Acknowledgments}
BM, TG and SG acknowledge Luxembourg National Research Fund (FNR) for supporting this work through the project TRICOLOR (INTER/NWO/20/15079143/TRICOLOR). The authors would like to thank Prof. Beatriz Noheda and Johanna van Gent  for valuable discussions on the results. 

\section*{Supporting Information}
The Supporting Information is available free of charge on the ACS Publications website and from the author. Polarization hysteresis of ZHO film, EDS and FFT diffractogram of HZO film, leakage and fatigue measurements of HO and HZO films.

\section*{Declaration of Interest}
The authors declare no competing interests.

\section*{Data Availability statement}
The raw data generated in this study have been deposited in the Zenodo repository under https://doi.org/10.5281/zenodo.14893377.

\clearpage
\printbibliography[heading=bibintoc,title={References}]

\end{document}

\maketitle

\clearpage
\section*{S1. ZHO film}

\begin{figure}[!ht]
        \centering
            \includegraphics[scale=0.6]{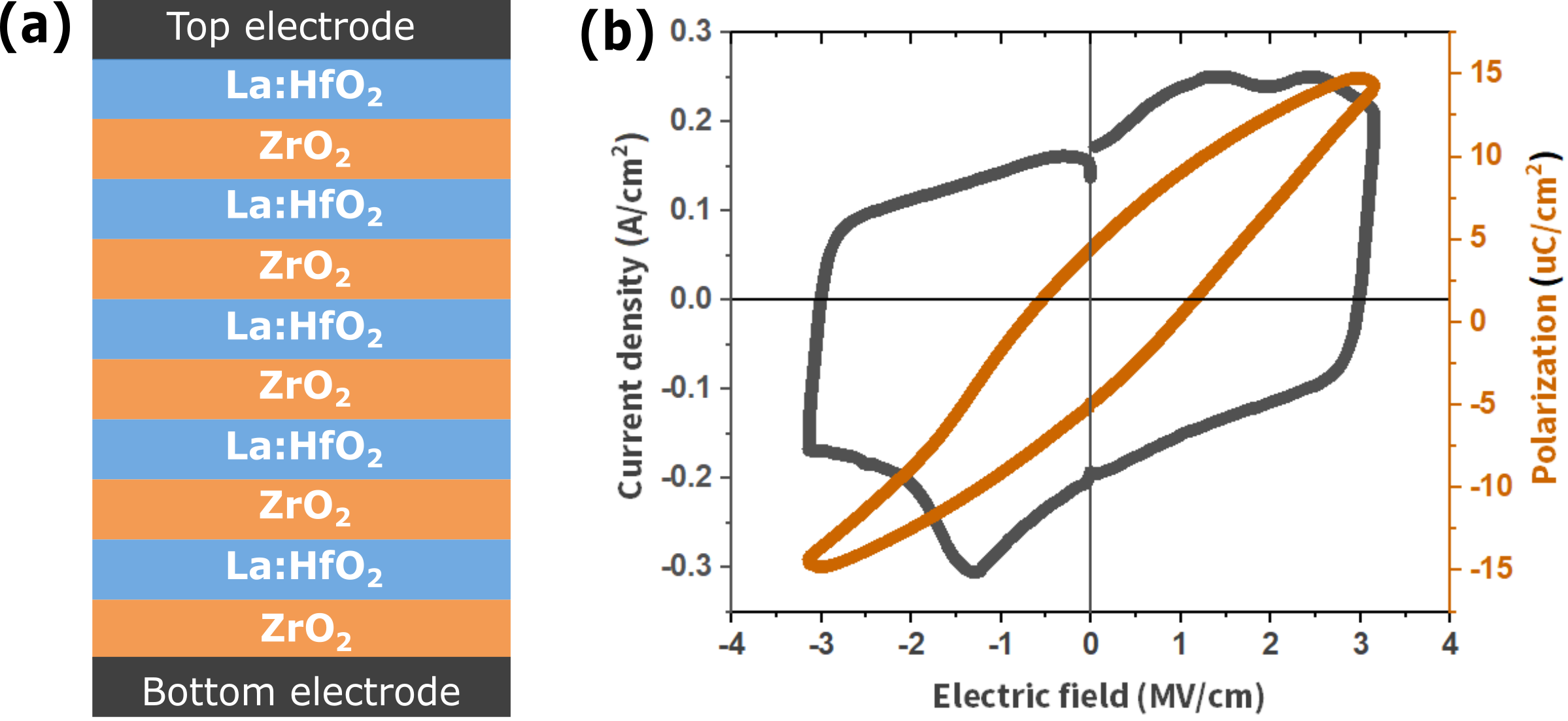}
            \caption{\footnotesize (a) Structural schematics, and (b) hysteresis loops of the ZHO film after treating with procedure 2.}
            \label{fig:ZHO}
    \end{figure}

    Figure \ref{fig:ZHO}(a) presents a structural schematic of the multilayer ZHO film, which consists of ten 5 nm-thick layers of alternately deposited ZrO\textsubscript{2} and La:HfO\textsubscript{2}, starting with ZrO\textsubscript{2} and ending with La:HfO\textsubscript{2}. Figure \ref{fig:ZHO}(b) presents the electrical characterization of the ZHO film. After treatment with procedures 1 and 2, the ZHO film hardly shows switching current.

\section*{S2. Energy-dispersive X-ray spectroscopy (EDS)}

\begin{figure}[!ht]
        \centering
            \includegraphics[scale=0.55]{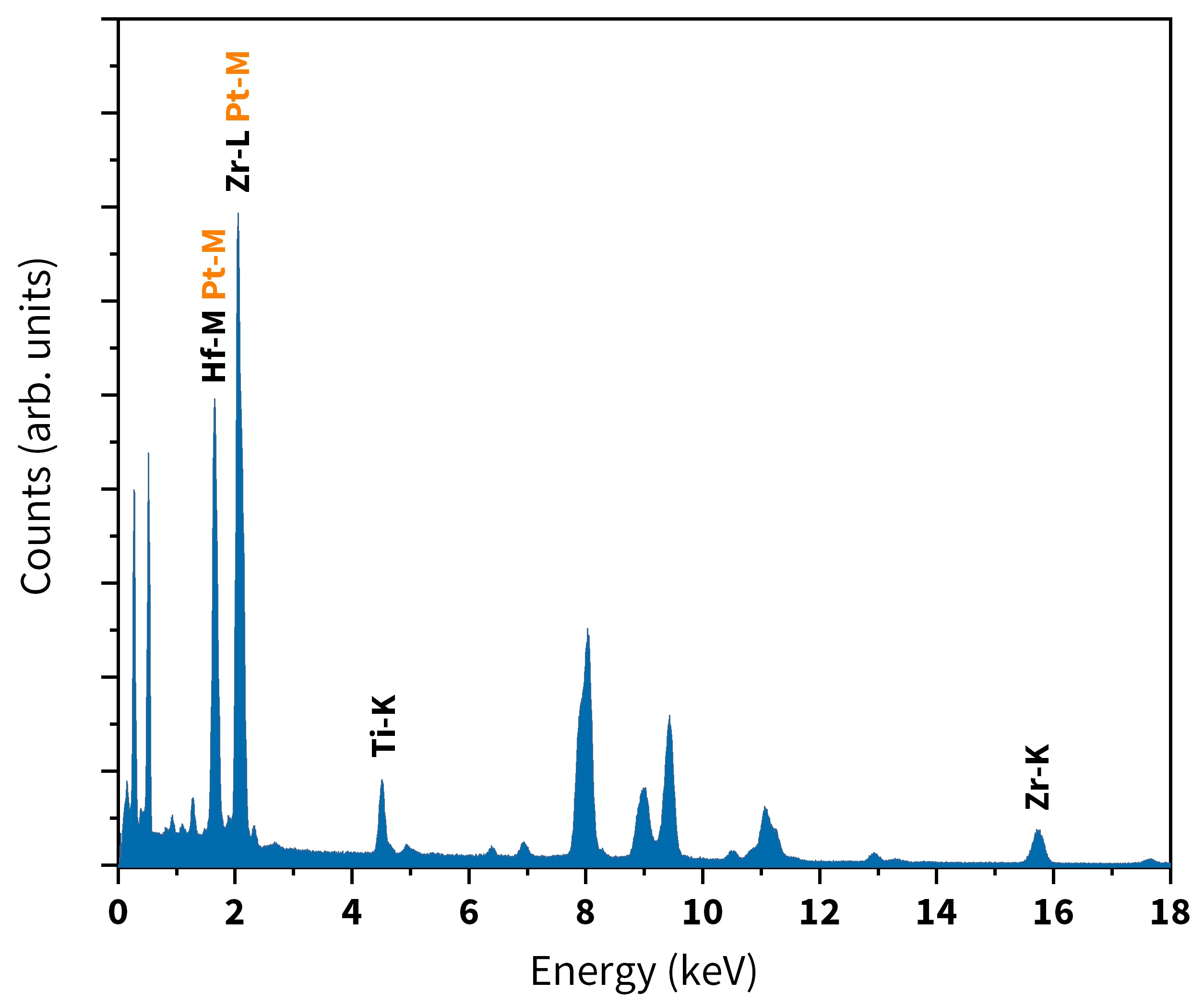}
            \caption{\footnotesize EDS spectrum of HZO sample cross-section.}
            \label{fig:suppl_eds}
    \end{figure}

\begin{figure}[!ht]
        \centering
            \includegraphics[scale=0.5]{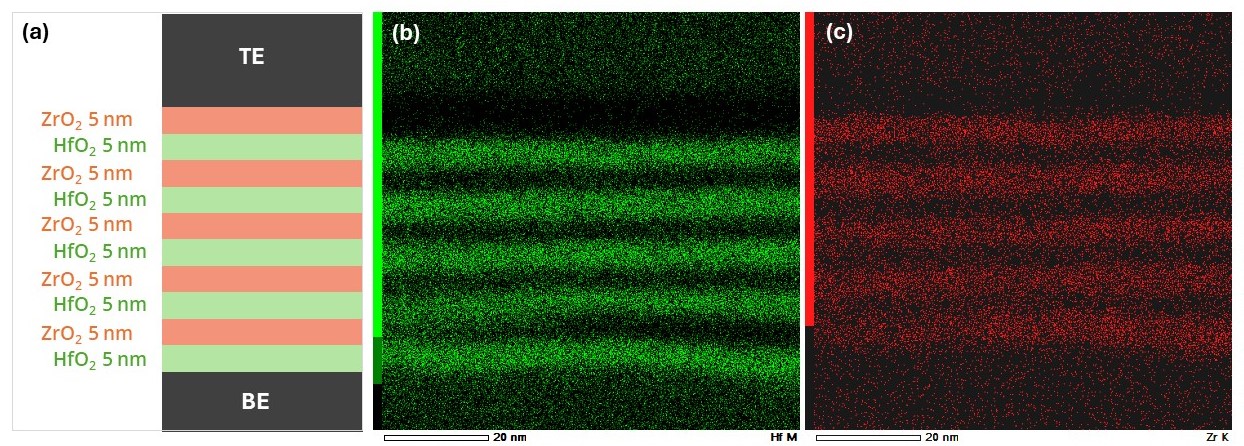}
            \caption{\footnotesize (a) Multilayer schematic. EDS mapping of the cross-section profile of the HZO multilayer film consisting of 
            (b) hafnia layers extracted from Hf-M and (c) zirconia layers extracted from Zr-K line profiles. TE: top electrode, BE: bottom electrode.}
            \label{fig:suppl_cs}
    \end{figure}

    The EDS spectrum in Figure \ref{fig:suppl_eds} was utilized for line profile analysis. The EDS mapping of the cross-section profile of the HZO multilayer film is presented in Figure \ref{fig:suppl_cs}. Note that the green regions observed both at the top and at the bottom of the Hf map (\ref{fig:suppl_cs} (b)) are an artifact due to the superimposition of the EDS Hf M and Pt M peaks. Similarly, the red regions visible both at the top and at the bottom of the Zr map (\ref{fig:suppl_cs}(c)) is an artifact due to the superimposition of EDS Zr K and Pt M peaks.

    \
    
    \clearpage

\section*{S3. Fast Fourier Transform (FFT)}
    \begin{figure}[!ht]
        \centering
            \includegraphics[scale=0.65]{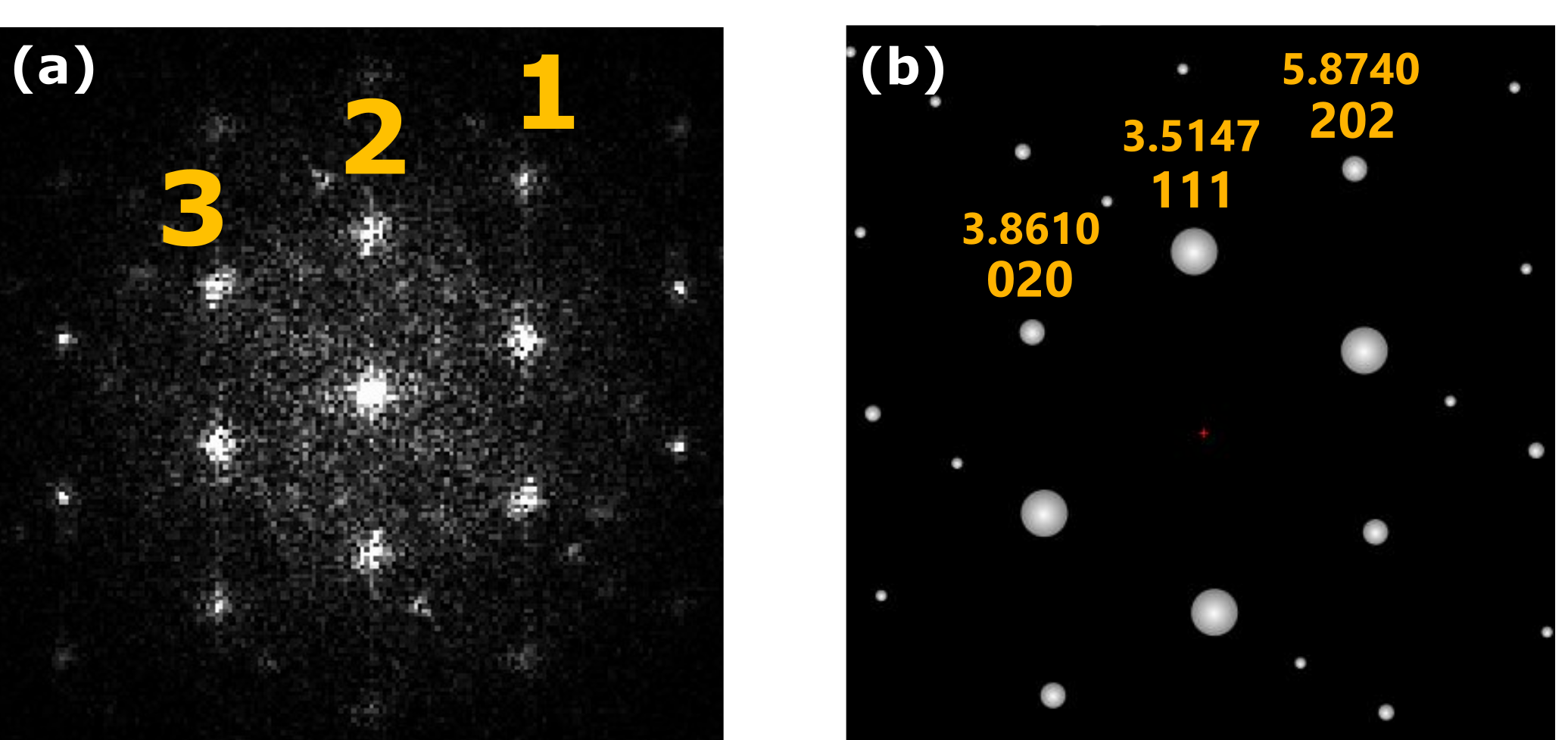}
            \caption{\footnotesize (a) FFT diffractogram of the sample. (b) Simulated diffraction pattern of monoclinic HfO\textsubscript{2} along [-101] zone axis (space group P2\textsubscript{1}c).}
            \label{fig:suppl_fft_fits}
    \end{figure}
    
    Figure \ref{fig:suppl_fft_fits} shows the FFT diffractogram of the sample alongside simulated diffractograms for the m-phase. The bright central spot of the bi-dimensionnal FFT is a consequence of the smooth, i.e. long-range, transitions as well as the presence of noise in the original image Figure  \ref{fig:suppl_fft_fits}(a). Indeed, the center of the FFT is where the low frequency components of the real space are stored. All other bright spots visible on Figure  \ref{fig:suppl_fft_fits}(a) are related to higher frequencies corresponding to short-range periodicities on the original image and their respective spacing are measured from the center of the FFT image. A comparison of the spacings is provided in Table \ref{table_a}. The two-spot ratio of the sample is more closely aligned with the o-phase and t-phase than with the m-phase. \\    
        \begin{table}[h]
        \begin{center}
        \small
        \caption{Comparison of FFT spot spacing (in nm\textsuperscript{-1}) between the HZO sample data and the simulated phases shown in \ref{fig:suppl_fft_fits}.}.\\
        \label{table_a}
        \begin{tabular}{c c c c c}
        \hline
         & HZO multilayer & o-phase & t-phase & m-phase\\
        \hline
        Spot 1 & 5.29 & 5.54 & 5.39 & 5.87 \\
        Spot 2 & 3.25 & 3.34 & 3.32 & 3.51 \\
        Spot 3 & 3.77 & 3.73 & 3.87 & 3.86 \\
        Ratio (spot 1/ spot 2) & 1.62 & 1.65 & 1.62 & 1.67 \\
        Ratio (spot 2/ spot 3) & 0.86 & 0.89 & 0.85 & 0.90 \\
        \hline
        \end{tabular}
        
        \end{center}
        \end{table}
    \clearpage

\section*{S4. Breakdown and fatigue measurements}
    
\begin{figure}[!ht]
        \centering
            \includegraphics[scale=0.60]{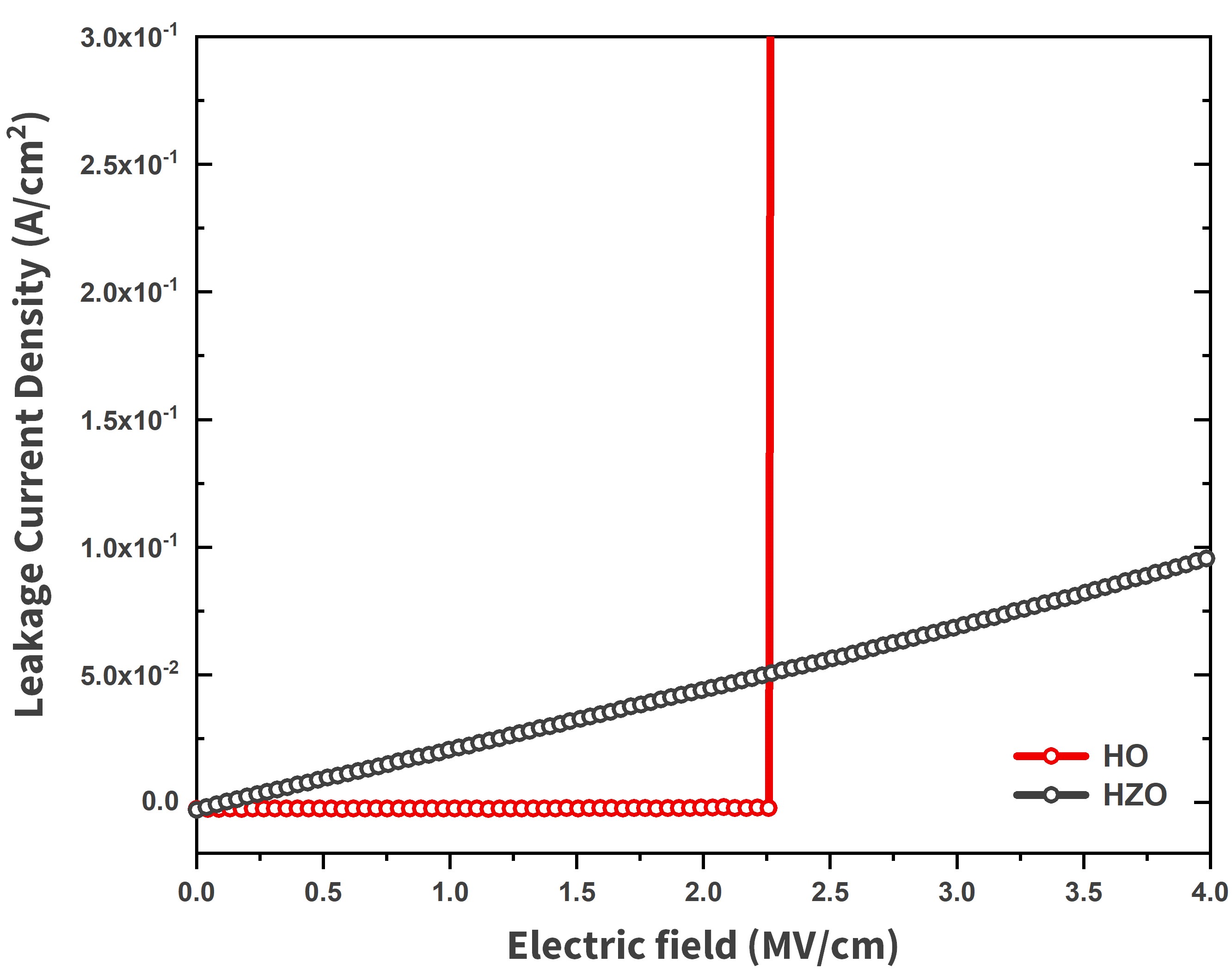}
            \caption{\footnotesize Breakdown measurements of the pristine HO and HZO films. The measurement was performed on the TF Analyzer 2000 (aixACCT) tool with a step of 200 mV with a waiting time of 100 s for each data point.}
            \label{fig:leakage}
    \end{figure}

    Repeated measurements were performed on both the samples. Figure \ref{fig:leakage} presents the breakdown measurements for one example of each type. The HO films consistently broke down at $\sim$2.3 MV/cm, while HZO sustained fields up to 4.0 MV/cm and beyond. Leakage current was below the measurement resolution for HO, while the HZO films showed an Ohmic behaviour with a resistivity ranging from 4 $\cdot$ 10\textsuperscript{7}  $\Omega$cm to 7 $\cdot$ 10\textsuperscript{7}  $\Omega$cm.
    
\begin{figure}[!ht]
        \centering
            \includegraphics[scale=0.55]{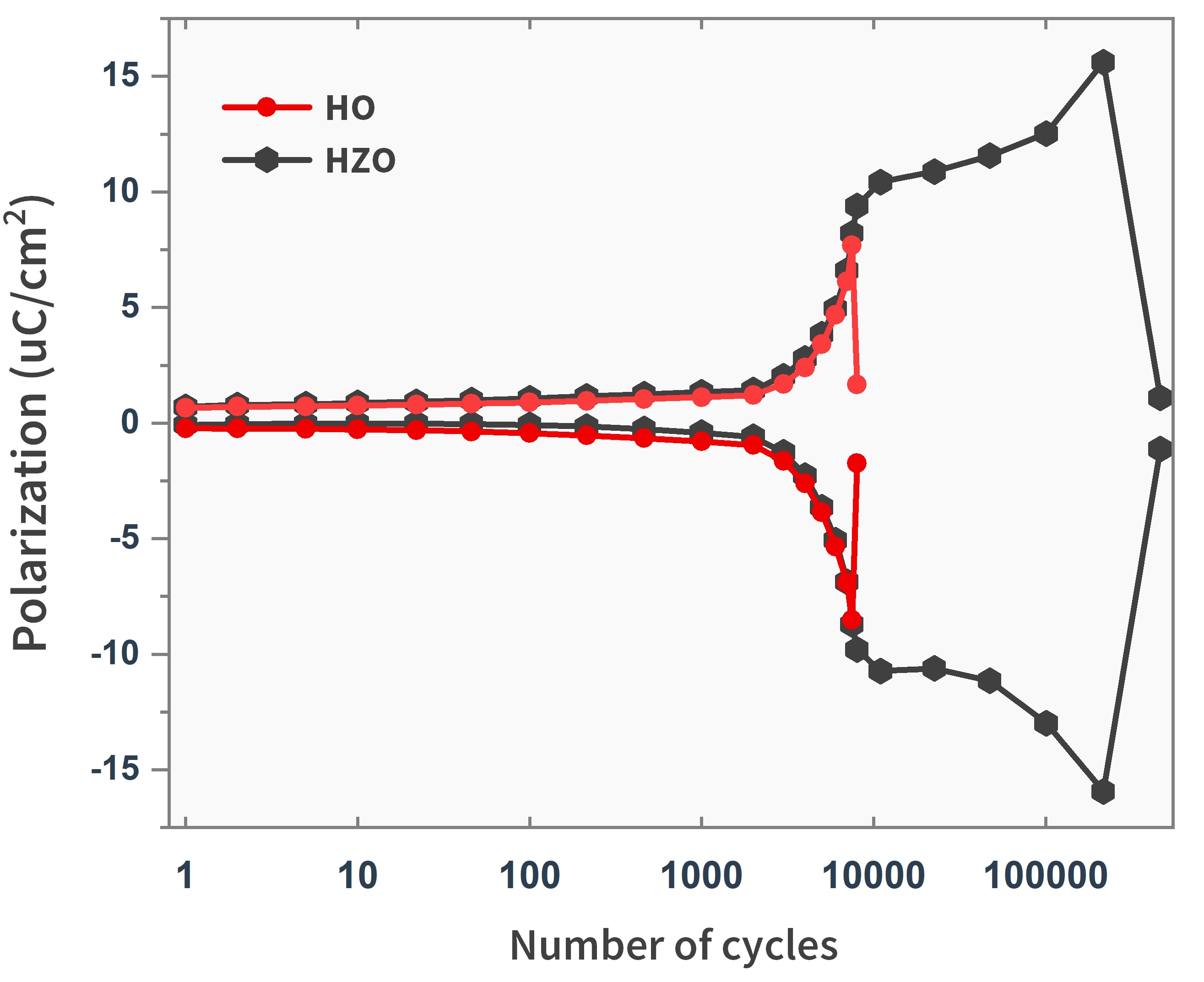}
            \caption{\footnotesize Fatigue measurements. The measurements were conducted by cycling at 3 kHz with gradually increasing the electric field from 0.8 MV/cm to 3 MV/cm (procedure 2).}
            \label{fig:fatigue}
    \end{figure}

    Fatigue measurements are shown in Figure \ref{fig:fatigue}. The measurements were conducted at 3 kHz with gradually increasing the electric field from 0.8 MV cm\textsuperscript{-1} to 3 MV cm\textsuperscript{-1} (procedure 2). The HO sample experienced breakdown after 7,000 cycles, whereas the HZO multilayer films endured over 100,000 cycles, lasting an order of magnitude longer than the HO film. However, significant leakage current was observed after 50,000 cycles.
    
    \vspace*{3in}